\documentclass[a4paper]{article}
\usepackage{graphicx}

\begin{document}

\title{
Discreteness-induced Stochastic Steady State in Reaction Diffusion Systems:
Self-consistent Analysis and Stochastic Simulations
}
\author{Yuichi Togashi and Kunihiko Kaneko}
\date{August 16, 2004}
\maketitle

\begin{abstract}
A self-consistent equation to derive a discreteness-induced
stochastic steady state is presented for reaction-diffusion systems.
For this formalism, we use 
the so-called Kuramoto length, a typical distance over which a molecule
diffuses in its lifetime, as was originally introduced
to determine if local fluctuations influence globally the whole system.
We show that this Kuramoto length is also relevant to determine
whether the discreteness of molecules is significant or not.
If the number of molecules of a certain species
within the Kuramoto length is small and discrete,
localization of some other chemicals is brought about, which can accelerate certain reactions.
When this acceleration influences the concentration of the original molecule species, 
it is shown that  a novel, stochastic steady state is induced
that does not appear in the continuum limit.
A theory to obtain and characterize this state is introduced, based on the
self-consistent equation for chemical concentrations.
This stochastic steady state is confirmed by
numerical simulations on a certain reaction model,
which agrees well with the theoretical estimation.
Formation and coexistence of domains with 
different stochastic states are also reported, which is maintained by the discreteness.
Relevance of our result to intracellular reactions is briefly discussed.
\end{abstract}

\section{Introduction}

Chemical reaction dynamics are often studied with the use of
rate equations for chemical concentrations.  For this approach,
the number of molecules is assumed to be large, which validates the
continuum description.  However,
in a biological system such as a cell, the number of molecules within
is sometimes rather small.  Then the
validity of continuum description by the rate equations is not evident.
This problem of smallness in molecule number is not restricted in biology.
Following recent advances in nanotechnology, reactions in 
a micro-reactor are studied experimentally, where the number of molecules 
in concern is quite small.  This is also true in some 
surface reaction of absorbed chemicals.

Here we are interested in the effect of such smallness in molecule number. Of course,
one straightforward consequence of the smallness in the number is the large fluctuations
in the concentration.  Indeed, the fluctuations around the continuous rate equation can
be discussed by stochastic differential equation \cite{vanKampen,KMK}.
State change by noise has been studied as noise-induced transitions \cite{NIP-book}, 
noise-induced
order \cite{Tsuda-Matsumoto}, stochastic resonance \cite{Parisi}, and so forth.
The use of stochastic differential equation, as well as its consequence, has been investigated
thoroughly.
If the number of molecules is much smaller and can reach 0, however, another effect 
of "smallness" is expected, that is the discreteness in the number.
Our concern in the present paper
is a drastic effect induced by such discreteness in the molecule number.

Previously we have discovered a transition of a chemical state,
induced by the discreteness in the molecule number, i.e., the
effect of the number of molecules $0$, $1$, $\cdots$ \cite{PRL}.  The transition is
termed as discreteness-induced transition (DIT).
This transition is not explained by the stochastic differential equation approach.  Rather,
discreteness, in conjunction with the stochastic effect, is essential.   In particular,
we studied a system with autocatalytic reaction in a well stirred container.
As the volume of the container decreases and the molecule number decreases, 
a transition to a novel state with symmetry breaking occurs, that does not appear
either in the continuous rate equation or in its Langevin version.
Here the transition occurs, when the number of molecule flow from environment to the reactor
is discrete, in the sense that it is less than one on the average, within the average reaction time.
Indeed, to the discreteness-induced transition,
relevant is not the molecule number itself but
the discreteness in the number of some molecular process (e.g., flow of molecule into the system)
within the average time scale of some other reaction process.

On the other hand, in a spatially extended system with reaction and diffusion, 
the total number of molecules (and molecular events) increases with the system size, and is
not small.  Instead,  
the number of molecules (or events), not in the total system but
within the size of an ``effective length'', is relevant to determine the discreteness effect.
Then, we need to answer what this effective length is.
In \cite{PRE}, we have proposed that the so-called Kuramoto length gives an answer to it.

Kuramoto length $\ell_K$ is defined as
the average length that a molecule diffuses within its lifetime, i.e.,
before it makes reaction with other molecules \cite{vanKampen,Kuramoto1,Kuramoto2}.
In the seminal papers \cite{Kuramoto1,Kuramoto2}, Kuramoto has shown that whether the total system size is larger 
than this length or not provides a condition to guarantee the use of
the reaction-diffusion equation.
When the system size (length) is smaller than $\ell_K$,
local fluctuations rapidly spread over the system.
Contrastingly, if the system size is much larger than $\ell_K$,
distant regions fluctuate independently, and the system is described by
local reaction process and diffusion, validating the use of
reaction-diffusion equation.

For example, consider the reaction
\[
A \stackrel{k}{\longrightarrow} X,\quad 2X \stackrel{k'}{\longrightarrow} B.
\]
If the concentration of chemical $A$ is set to be constant,
the chemical $X$ is produced at the constant rate $k$,
while it decays with the reaction $2X \rightarrow B$ at the rate $k'$.
The average concentration of $X$ at the steady state is
$\langle X \rangle = \sqrt{kA/2k'}$,
where $A$ is the concentration of the chemical $A$.
Thus the average lifetime of $X$ at the steady state is estimated to be
$\tau = 1 / (2k'\langle X \rangle) = 1 / \sqrt{2kk'A}$.
If $X$ molecules diffuse at the diffusion constant $D$ in
one-dimensional space,
the typical length over which an $X$ molecule diffuses in its lifetime
is estimated to be
\begin{equation}
\ell_K = \sqrt{2D\tau},
\label{eqn:kuramoto}
\end{equation}
which gives the Kuramoto length.

In these works, it is assumed that the average distance between
molecules is much smaller than $\ell_K$, and there is a large number of molecules within the
region of the length $\ell_K$.  Thus
the concentration of the chemical $X$ can be
regarded as a continuous variable.  Hence the continuum description is valid.
However, if the average distance
between molecules is comparable to or larger than $\ell_K$, local
discreteness of molecules may not be negligible.  

For example, consider a chemical species $X_j$, whose Kuramoto length is given by
$\ell _j$.  Then we consider discreteness of molecule species
$X_{i}$ that produces this chemical $X_j$, i.e., the case that  average number of $X_{i}$
is less than 1 within the area of the Kuramoto length $\ell _j$. 
With this setting, molecules $X_{j}$, produced by $X_{i}$ molecules, will be localized around them,
as the average distance between  $X_{i}$ molecules is larger than 
the Kuramoto length of $X_j$.  Then, 
this localization
of the  chemical may drastically alter the total rate of the reactions,
if reactions with 2nd or higher order of $X_{j}$ are involved, as will be shown later. 
In the present paper, following \cite{PRE}, we pursue the possibility that discreteness of
some molecules within Kuramoto length of some other molecules
may drastically change the steady state of the system, as in DIT previously studied.

In section 2, we discuss a general condition for the amplification of some reaction by
such discreteness.  Then by introducing a self-consistent equation for the rate of this
amplification, we demonstrate the existence of
stable stochastic steady state (SSS), that never appears in the continuum description.
In section 3, we numerically study a specific chemical reaction model with three components,
to show the validity of this self-consistent theory for SSS.  In section 4, domain formation with
this SSS is presented, as a novel possibility for pattern formation in reaction-diffusion
system.
Discussion is given in section 5, with possible applications to biological problems.

\section{Steady state induced by discreteness of molecule, with amplification of 
some reaction: self-consistent analysis}

Consider a reaction system consisting of several molecule species $X_m$ ($m=1,...,k$), with
chemical reaction and diffusion.
The system can involve catalytic reactions of higher-order catalysis or autocatalysis.
Some other molecules (e.g., resource chemicals)
are supplied externally, involved in the reaction among $X_m$,
so that the nonequilibrium condition is sustained.  So far the system in concern is
rather general chemical reaction system with diffusion.

Now, we take a pair of
molecule species, $X_i$ and $X_j$, where $X_j$ is produced by $X_i$, and study
how the discreteness of the molecule $X_i$ can alter the steady state from the continuum limit case.
To discuss the discreteness effect,
we consider the case that the molecule $X_j$ is localized around the molecule $X_i$.
(Recall the molecule $X_j$ is produced by $X_i$.)  In order for this localization to work, the
average length that the molecule $X_j$ travels within its lifetime should be
smaller than the average distance of the molecules $X_i$.  In other words, the average number of
the molecules $X_i$ within the domain of the Kuramoto length $\ell_j$ is less than 1.
(see Fig. 1).  Here, the lifetime of the molecule $X_j$ is determined by the collision
with some other molecule species whose density is not low.  Hence, the Kuramoto length
of the $X_j$ molecule, determined as in \S 1, is given independently of
the concentrations of the molecules $X_i$ and $X_j$.

Now, to alter drastically the steady state by discreteness, the localization of $X_j$ molecule
has to change concentrations of some other molecules, as compared with the
case of homogeneous distribution of $X_j$.  This is possible
if there is a higher-order reaction such as $mX_j+X_q \rightarrow X_p$, because the
probability of such reaction is amplified by localization of $X_j$ molecules in space.
To compute this acceleration, we calculate
the average of $c_{j}^{m}$, where $c_{j}$ is the concentration of $X_{j}$,
and compute the degree of
amplification $\alpha$ from that for
the homogeneously distributed  case.
In the calculation, we assume that $X_{j}$ is localized around the $X_{i}$ molecules, with a width of
$\ell_j$, the Kuramoto Length of $X_{j}$ (which is shorter than the average
distance between $X_{i}$ molecules).

Assuming that the distribution of $X_{j}$ is represented by
the continuous concentration $c_{j}(\vec{x})$,
$\alpha$ can be expressed as
\begin{equation}
\alpha = \frac{\langle c_{j}^{m} \rangle}{\langle c_{j} \rangle^{m}} = \frac{V^{-1}\int c_{j}^{m} d\vec{x}}{\left(V^{-1} \int c_{j} d\vec{x}\right)^{m}},
\label{eqn:alpha0}
\end{equation}
where $V$ is the size (volume) of the system.

For simplicity, we assume that $X_{i}$ is randomly distributed over
$d$-dimensional space,
and the distribution of $X_{j}$ is given by a $d$-dimensional Gaussian distribution
with a standard deviation $\ell_{j}$ around each $X_{i}$ molecule,
such as
\[
\rho_{k}(\vec{x}) = \frac{1}{(\sqrt{2\pi}\ell_{j})^{d}} \exp\left(- \frac{|\vec{x} - \vec{x_{k}}|^{2}}{2 \ell_{j}^{2}}\right),
\]
where $\vec{x_{k}}$ is the position of each $X_{i}$ molecule.
Now $c_{j}(\vec{x})$ is the sum of $\rho_{k}(\vec{x})$;
thus, $\langle c_{j} \rangle = c_{i}$ since $\int \rho_{k} d\vec{x} = 1$.
For the case with $m=2$ and sufficiently large $V$,
\begin{eqnarray}
\langle c_{j}^{2} \rangle &=& \left\langle (\textstyle\sum \rho_{k})^{2} \right\rangle\nonumber\\
 &=& \left( \textstyle\sum \langle \rho_{k} \rangle \right)^{2} + \textstyle\sum \langle \rho_{k}^{2} \rangle
= \langle c_{j} \rangle^{2} + (2\sqrt{\pi}\ell_{j})^{-d} \langle c_{j} \rangle ,\nonumber
\end{eqnarray}
since $X_{i}$ is randomly distributed.
With eq. (\ref{eqn:alpha0}),
\begin{eqnarray}
\alpha &=& \left(\langle c_{j} \rangle^{2} + (2\sqrt{\pi}\ell_{j})^{-d} \langle c_{j} \rangle\right) / \langle c_{j} \rangle^{2}\nonumber\\
&=& 1 + (2\sqrt{\pi}\ell_{j})^{-d} \langle c_{j}\rangle^{-1}
= 1 + (2\sqrt{\pi})^{-d} c_{i}^{-1} \ell_{j}^{-d}\nonumber
\end{eqnarray}
Thus, we obtain the acceleration factor
\begin{equation}
\alpha = 1 + \frac{1}{(2 \sqrt{\pi})^{d} \ c_{i}\ell_{j}^{d}}.
\label{eqn:alpha1}
\end{equation}
In the same manner, for $m=3$,
\begin{equation}
\alpha = 1 + \frac{3}{(2 \sqrt{\pi})^{d} \ c_{i}\ell_{j}^{d}} + \frac{1}{(2\sqrt{3}\pi)^{d} \ (c_{i}\ell_{j}^{d})^{2}},
\end{equation}
and generally, for $c_{i}\ell_{j}^{d} \ll 1$,
\begin{equation}
\alpha \approx m^{-d/2} \left( (2\pi)^{d/2} c_{i} \ell_{j}^{d} \right)^{1-m}.
\end{equation}
As shown, the reaction can be drastically amplified as the number of $X_i$ molecules
within a volume of the Kuramoto length ($c_{i}\ell_{j}^{d}$) is much smaller than 1.

So far we have shown that for reaction system involving the process from $X_i$ to $X_j$, the
discreteness can alter the concentration of some chemicals drastically,
if (1) the density of $X_i$ molecule is so low that the number is discrete within the size of
the Kuramoto length of $X_j$ molecule and (2) there is a high order (higher than linear) 
reaction with regards to $X_j$.

Next, to confirm that this acceleration of reaction alters the steady state from the continuum case,
we need to check if  the condition for the discreteness is sustained under the
above amplification of concentration of some chemicals, as a steady state solution.
Hence we study some feedback from the concentration of $X_p$ to
$X_i$ that is generated by some reaction path(s).  If $X_i$ is produced or
catalyzed by $X_p$, the concentration of $X_i$ depends on that
of $X_p$, $c_p$.
With such feedback, the change of concentration $c_i$ is given by some function $F(c_j,c_p)$, while
the change of the concentration $c_j$ depends on $c_i$, and
is given by some function as $G(c_i)$.
Since $c_p$ is a function of $\alpha(c_i)$, $F(c_j,c_p)$ is rewritten as $\hat{F}(c_j,\alpha(c_i))$.
Hence the concentrations of $X_i$ and $X_j$  molecules must satisfy
\begin{equation}
dc_i/dt=\hat{F}(c_j,\alpha(c_i));\ dc_j/dt=G(c_i)
\label{eqn:cicj}
\end{equation}
($F$ and $G$ may have dependence on other concentrations or reaction rates,
e.g., $G(c_i)$ can also depend on $c_j$ or $\alpha$).
The steady state solution is obtained by setting the right hand of these equations as 0.

Note that the solution with $\alpha=1$ corresponds to the continuum case,
given by the  standard rate equation.  For some case, this is the only solution for the 
concentrations.
For some other cases, however, there is some other solution(s)
with $\alpha >1$.  This is a solution with the amplification
by localization of molecules due to the discreteness of the $X_i$ molecule.
If the concentration of $c_i$ obtained from this solution
satisfies $c_i \ell_j^{d}<1$,
this discreteness-induced solution is self-consistent.
Furthermore, the stability of this solution is computed by linearizing the solution
around this fixed point.  If this solution is linearly stable, stability
of this novel steady solution is assured, which does not exist in the continuum description
(or in its Langevin equation version). We call the state represented by
this solution as stochastic steady state (SSS), as it is sustained 
stochastically through discreteness in molecule numbers.
We will show an explicit example of this SSS in the next section.

\subsection*{Self-consistent solution involving the change of Kuramoto-length}

So far we have assumed that the Kuramoto-length $\ell _j$ of the $X_j$ molecule
is constant.  This is true as long as the concentration of the molecule 
relevant to the decomposition or transformation of the $X_j$  molecule is
constant.  However, if the concentration of the chemical that is relevant to the 
determination of $\ell_j$ depends on the concentration of either $X_i$, $X_j$, or
$X_p$, the Kuramoto length, as well as $\alpha$, depends on it. Accordingly, in eq. (\ref{eqn:cicj}),
we need to regard $\ell_j$ in $\alpha$ as a variable that depends on
either $c_i$, $c_j$ or $\alpha$.   With the inclusion of the dependence, we again
obtain a self-consistent solution, to get the concentrations $c_i$ and $c_j$ 
(and accordingly $\alpha$ and $\ell_j$).  If there is a stable solution with
$\alpha>1$ and $c_i \ell_j^{d}<1$, then we get a SSS as a self-consistent
solution both on $\alpha$ and $\ell_j$.  We will discuss a related example in \S4, where
two solutions with $\alpha=1$ and $\alpha>1$ coexist in space, and form a domain structure.

\subsection*{Combination of several processes}

So far we have discussed a simple case of discreteness-induced state.
The discussion with the use of amplification factor $\alpha$, however, is
generalized to include temporally or spatially dependent solutions of $c_i$ and $c_j$,
with temporal (or spatial) dependence of $\alpha$.  
This solution represents an average behavior longer than the time scale for
stochastic collisions or 
longer scale than $\ell_j$.
With this extension,
we can discuss discreteness-induced rhythm or pattern, that is stochastically sustained.

Such spatiotemporal dynamics can often appear
in a reaction network of several molecules, with two or more
pairs of discreteness in number.  For example, consider reactions
$X_{i1} \rightarrow(produces)\rightarrow X_{j1}$, $X_{j1} \rightarrow(produces\ with\ high\ order\ reaction) \rightarrow X_{p1}$, $X_{p1} \rightarrow(produces)\rightarrow X_{i2}$; and
$X_{i2} \rightarrow(produces)\rightarrow X_{j2}$, $X_{j2} \rightarrow(produces\ with\ high\ order\ reaction) \rightarrow X_{p2}$, $X_{p2} \rightarrow(produces)\rightarrow X_{i1}$
(see Fig. \ref{fig:sp1-cascade-loop}), where we assume that the density $c_{i1}$, $c_{i2}$, of $X_{i1}$ and $X_{i2}$ 
molecules are so low that $c_{i1}\ell_{j1}^{d}<1$ and $c_{i2}\ell_{j2}^{d}<1$, respectively
for the Kuramoto lengths of $X_{j1}$ and $X_{j2}$. Then,  following
the scheme we discussed, we get a coupled equation for the concentrations of
$c_{i1}$, $c_{j1}$, $c_{i2}$, $c_{j2}$ with two amplification factors
$\alpha_1$ and $\alpha_2$.  In general, there may be a
time- or space- dependent solution
(by including diffusion term with  much longer spatial scale),  
that leads to a novel stochastic pattern or rhythm.
Explicit examples for such case will be discussed in future.


\begin{figure}
\begin{center}
\includegraphics[width=80mm]{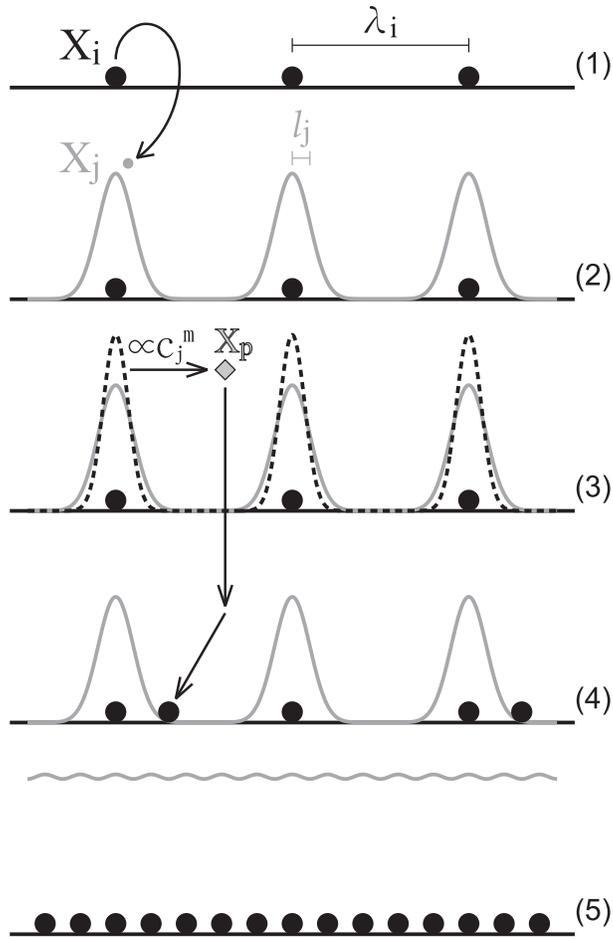}
\end{center}
\caption{Schematic representation for the mechanism to produce SSS.
Assume that (1) the chemical $X_{j}$ is produced by $X_{i}$,
and (2) $\ell_{j}$, the Kuramoto length of $X_{j}$, is shorter than
$\lambda_{i} (= c_{i}^{-1})$, the average distance of $X_{i}$ molecules
(i.e., $c_{i}\ell_{j}^{d} < 1$).
Then, $X_{j}$ is localized around $X_{i}$ molecules.
(3) If there is $m$-th order reaction of $X_{j}$ ($m>1$),
the rate of the reaction is proportional to $c_{j}^{m}$;
hence, the reaction is accelerated.
(4) Additionally, if the reaction promotes (directly or indirectly)
the production of $X_{i}$, the acceleration of the reaction may cause
increase of $c_{i}$.
(5) On the other hand, if $c_{i}$ is high ($c_{i}\ell_{j}^{d} \approx 1$ or $>1$),
$X_{j}$ is almost uniformly distributed;
thus, the acceleration is weak, and the production of $X_{i}$ is degraded.
Hence, there is a stable steady state of $c_{i}$ at an intermediate value.}
\label{fig-sp1-sss}
\end{figure}

\begin{figure}
\begin{center}
\includegraphics[width=80mm]{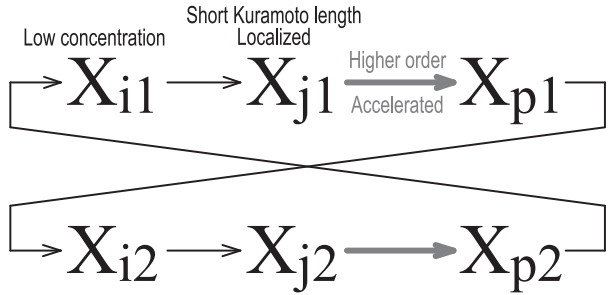}
\end{center}
\caption{Schematic diagram of an example of reaction cascade.}
\label{fig:sp1-cascade-loop}
\end{figure}

\section{Specific example of Stochastic Steady State}

To confirm our theoretical estimation for SSS,
we have adopted a simple model and carried out
stochastic particle simulations.
Here we consider a simple one-dimensional
reaction-diffusion system
with three chemicals ($X_{1}$, $X_{2}$, and $X_{3}$) and
four reactions:
\begin{eqnarray*}
X_{2} + X_{3} &\stackrel{k_{1}}{\longrightarrow}& X_{2} + X_{1},\\
X_{3} + X_{1} &\stackrel{k_{2}}{\longrightarrow}& 2 X_{3},\\
2 X_{2} &\stackrel{k_{3}}{\longrightarrow}& X_{2} + X_{1},\\
2 X_{1} &\stackrel{k_{4}}{\longrightarrow}& X_{1} + X_{2}.
\end{eqnarray*}
Here, we assume $k_{1}, k_{2} \gg k_{3} > k_{4}$.
We take $k_{1}=k_{2}=b r$, $k_{3}=a r$, and $k_{4}=r$ ($r > 0$, $1 < a \ll b$)
for further discussion.
We assume that the system is closed with regards to the molecules $X_m (m=1,2,3)$.
Thus, $N$, the total number of molecules (or $c$, the total concentration),
is conserved.

In the continuum limit, each $c_{i}(t,x)$,
the concentration of $X_{i}$ at time $t$ and position $x$,
obeys the following reaction-diffusion equation:
\begin{eqnarray}
\frac{\partial c_{1}}{\partial t} & = & -b r(c_{1}-c_{2})c_{3} - r(c_{1}^{2} - a c_{2}^{2}) + D_{1} \frac{\partial^{2} c_{1}}{\partial x^{2}},\ \label{eqn:rd1}\\
\frac{\partial c_{2}}{\partial t} & = & r(c_{1}^{2} - a c_{2}^{2}) + D_{2} \frac{\partial^{2} c_{2}}{\partial x^{2}}, \label{eqn:rd2}\\
\frac{\partial c_{3}}{\partial t} & = & b r(c_{1}-c_{2})c_{3} + D_{3} \frac{\partial^{2} c_{3}}{\partial x^{2}}, \label{eqn:rd3}
\end{eqnarray}
where $D_{i}$ is the diffusion constant of $X_{i}$.
For simplicity, we assume $D_{i} = D$ for all $i$.
This reaction-diffusion equation has homogeneous fixed point solutions
with $(c_{1}, c_{2}, c_{3}) = (0, 0, c)$, $(\sqrt{a}c/(\sqrt{a} + 1), c/(\sqrt{a}
+ 1), 0)$ for all $x$.  By linear stability analysis, it is straightforward to show that
only the former is stable.  Starting from any initial conditions,
the partial-differential equation system simply converges to this stable fixed point.

In this system, the chemical $X_{1}$ is produced by
$X_{2}$ molecules. (In the notation of \S 2, $i=2$ and $j=1$.)
If $\ell_{1}$, the Kuramoto length of $X_{1}$, is shorter than the average
distance between $X_{2}$ molecules,
$X_{1}$ is localized around the $X_{2}$ molecules, as discussed in \S 2.
Then, the reaction $2 X_{1} \rightarrow X_{1} + X_{2}$,
which is at second order of $X_{1}$,
is accelerated.
Using eq. (\ref{eqn:alpha1}) in \S 2, we obtain the acceleration factor
\begin{equation}
\alpha = 1 + \frac{1}{2 \sqrt{\pi} c_{2} \ell_{1}}.
\label{eqn:alpha1a}
\end{equation}
On the other hand, the lifetime of $X_{2}$ is so long
that $X_{2}$ is not localized.
Thus, the reaction $2 X_{2} \rightarrow X_{2} + X_{1}$ is
not accelerated.

Now we study the self-consistent solution from $c_1$ and $c_2$, following the argument of \S 2.
(By a path from $X_1$ to $X_2$,
there is a direct  feedback to $i$, i.e., $p=2(=i)$ in the notation of \S 2.)
Here, we consider the case where $N_{1}$, $N_{2} \ll N_{3}$,
so that $c_{3} \approx c$.
Then, the average lifetime of $X_{1}$ is about $1/(brc_{3}) \approx 1/(brc)$;
we assume that $\ell_{1} = \sqrt{2D/(brc)}$ for further discussion.
When $N_{1}$, $N_{2} \ll N_{3}$, the two reactions
$X_{2} + X_{3} \rightarrow X_{2} + X_{1}$ and
$X_{3} + X_{1} \rightarrow 2 X_{3}$ are much faster than the other two
and maintain $N_{1} \approx N_{2}$.
Then, the rates of the other reactions satisfy
\begin{equation}
\frac{\textrm{The rate of }\ (2 X_{1} \rightarrow X_{1} + X_{2})}{\textrm{The rate of }\ (2 X_{2} \rightarrow X_{2} + X_{1})}
\approx \frac{\alpha k_{4} N_{1}^{2}}{k_{3} N_{2}^2}
\approx \frac{\alpha}{a}.
\label{eqn:r34ratio}
\end{equation}
When $\alpha = a$, the two reactions are balanced,
which leads to a novel fixed point.
Assuming $c_{1}, c_{2} \ll c_{3}$ and $c_{3} = c$, and
following eq. (\ref{eqn:alpha1a}), we obtain the condition
for the balance
\begin{equation}
c_{1} = c_{2} = \frac{1}{2\sqrt{\pi} (a-1) \ell_{1}} (= c_{s}).
\label{eqn:fixedpoint}
\end{equation}

Subsequently, we investigate the stability of this fixed point.
For $c_{1} = c_{s} + \delta c_{1}$ and $c_{2} = c_{s} + \delta c_{2}$,
we obtain
\begin{equation}
\alpha = 1 + \frac{(a - 1) c_{s}}{c_{2}}
= a - \frac{a - 1}{c_{s}}\delta c_{2} + o(\delta c_{2})
\label{eqn:alphafp}
\end{equation}
from eqs. (\ref{eqn:alpha1a}) and (\ref{eqn:fixedpoint}).
We take into account the acceleration factor
in the reaction-diffusion equation,
we obtain
\begin{eqnarray}
\left(
\begin{array}{c}
\dot{c_{1}}\\
\dot{c_{2}}
\end{array}
\right)
& = &
r \left(
\begin{array}{cc}
- 2a c_{s} - bc & (3a - 1) c_{s} + bc\\
2a c_{s} & - (3a - 1) c_{s}
\end{array}
\right)
\left(
\begin{array}{c}
\delta c_{1}\\
\delta c_{2}
\end{array}
\right) \nonumber
\\
 & & \ + o(\delta c_{1}, \delta c_{2}).
\label{eqn:fplinear}
\end{eqnarray}
For any $a$, $b > 1$, this Jacobi matrix has two negative eigenvalues,
implying that the fixed point is stable\footnote{It is also stable 
against spatially inhomogeneous perturbations for any $D > 0$.}.

In the simulations, we have found that
the system converges to the novel fixed point.  In fact,
we have measured $c_{2}$ at the fixed point
for a certain $\alpha = a$ numerically.
Figure \ref{fig:sp1-1-accel} shows the relation between
$1 / c_{2} \ell_{1}$ and $\alpha$,
which agrees rather well with the theoretical estimation in eq. (\ref{eqn:alpha1a}).

In summary, we demonstrated numerically that the discreteness of molecules
yields a novel stochastic steady state
in a reaction-diffusion system, in agreement with
the theoretical estimation.

\begin{figure}
\begin{center}
\includegraphics[width=80mm]{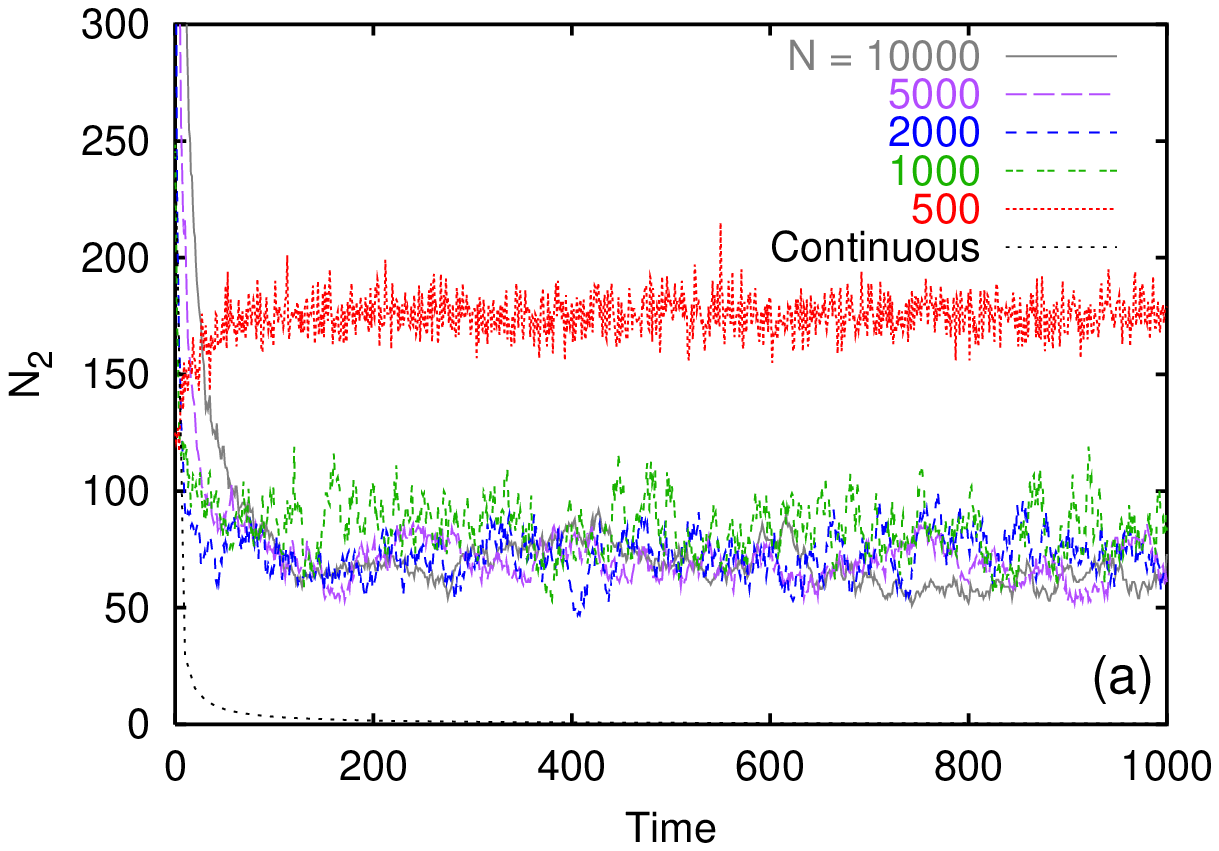}
\includegraphics[width=80mm]{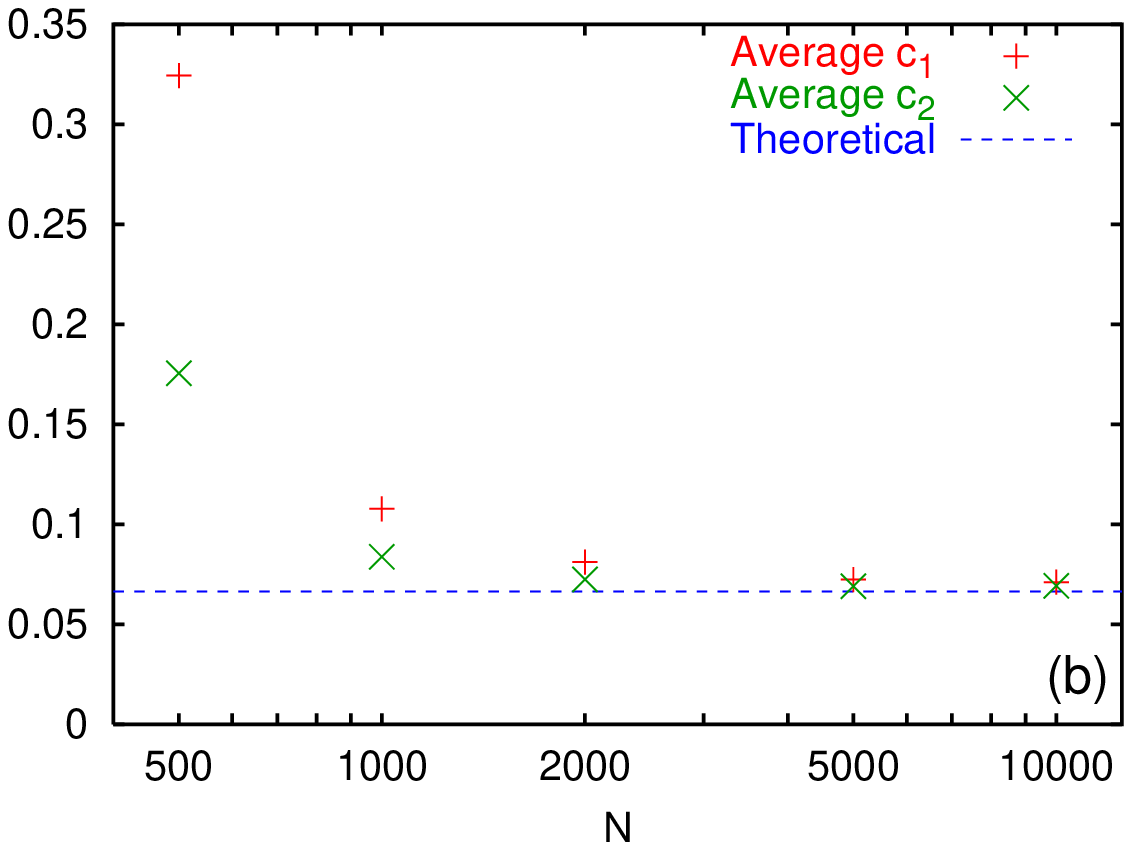}
\end{center}
\caption{(a) Time series of $N_{2}$ and (b) the average concentrations $c_{1}$ and $c_{2}$ for
several values of $N$.
$a=4$, $b=100$, $D=100$, and $L_{x}=1000$.
We fix $rc=1$ (i.e., $r=1000/N$), so that $\ell_{1}=\sqrt{2D/(brc)}=\sqrt{2}$.
For $N\ge 1000$ cases, $N_{2}$ converges to the stochastic fixed point.
There, $c_{2}$ should be $1/(2\sqrt{\pi}(a-1)\ell_{1}) = 1/6\sqrt{2\pi} \approx 0.066$ (i.e.,
$N_{2} = c_{2}L_{x} \approx 66$) for all cases, which agrees well with the simulation.
When $N=500$, the system reaches the unstable fixed point $(2c/3, c/3, 0)$.
With the reaction-diffusion equation, $c_{1}$ and $c_{2}$ rapidly
converge to $0$ (the dotted line ``continuous'' in (a) shows $c_{2}$ for $c=1$).}
\label{fig:ts-avec12}
\end{figure}

\begin{figure}
\begin{center}
\includegraphics[width=80mm]{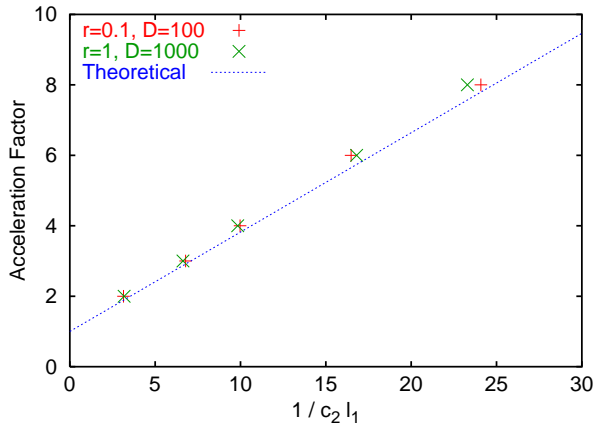}
\end{center}
\caption{
The acceleration factor $\alpha = a$,
plotted against $1 / c_{2} \ell_{1}$.
$b=100$, $N=1000$, $L_{x}=1000$, sampled over
$5000$ ($r=1$) or $50000$ ($r=0.1$) time units.
We measure the relation from average $c_{2}$ in simulations.
The result agrees with the theoretical estimation
$\alpha = 1 + \frac{1}{2\sqrt{\pi} c_{2} \ell_{1}}$ well.}
\label{fig:sp1-1-accel}
\end{figure}

\section{Coexistence of Domains with different Kuramoto Lengths}

In the example of the previous section, the spatial homogeneity is assumed
at a coarse-grained level, and indeed, this  homogeneous state was stable.
However, due to fluctuations,  some spatial inhomogeneity exists in SSS, and
a domain that is deviated from SSS may be produced.
Even if this deviated state is unstable in the continuum limit,
it may be preserved over a long time, if the concentration of
the molecule to destabilize it is so low that its discreteness
is essential.  If the average time to produce this deviated domain from SSS and
the lifetime of the state is balanced, the two regions,
SSS and the deviated state with different concentrations of molecules and
Kuramoto lengths, may coexist.
We give a simple example for it here.

\begin{figure}
\begin{center}
\includegraphics[width=80mm]{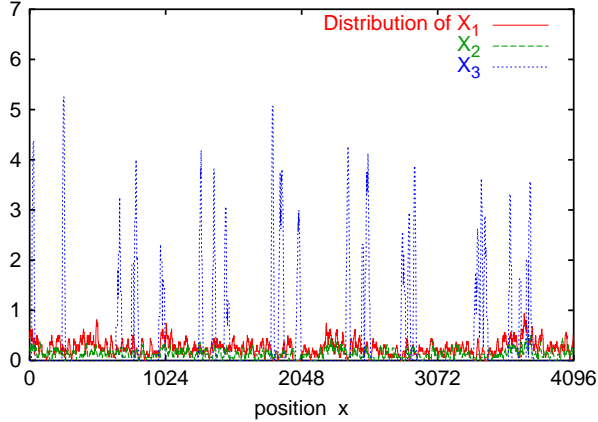}
\end{center}
\caption{The distribution of each chemical $X_{i}$ (a snapshot at $t=1000$).
$r=1.28$, $a=4$, $b=100$, $D=1$, $N=3200$, and $L_{x}=4096$.
(For these parameters, $\ell_{1B} \approx 17$ and $\ell_{3B} \approx 0.24$.)
Plotted distribution is obtained by averaging the molecule snapshot pattern,
with the bin size $\Delta x = 16$.
There appear spots of $X_{3}$, in which the stochastic steady state
with $\alpha > 1$ is realized.
There are very few $X_{3}$ molecules between the spots,
where the system stays around the unstable fixed point $(2c/3, c/3, 0)$.}
\label{fig:sp2-spa-3n2-nr4k-d1}
\end{figure}

Again we consider the same reactions as the preceding section.
Now, we assume that the diffusion of $X_{3}$ is slower than
the others,
and set $(D_{1}, D_{2}, D_{3}) = (100D, 100D, D)$ ($D > 0$).
In this model, there are two fixed-point states:
one is the  stochastic steady state mentioned above (which we call State A);
the other is the unstable fixed point with $c_{3}=0$ (State B), besides the
stable fixed point in the continuum limit (which does not appear here;
see Table \ref{tbl:fp} for the stability of the states mentioned here).
In Fig. \ref{fig:sp2-spa-3n2-nr4k-d1},
we give an example of snapshot pattern of the model.  In the figure,
except several spots with large $c_3$ that corresponds to the state A,
most other regions fall onto the state B that should be
unstable in the continuum limit.  Indeed, this pattern is not transient, and
the fraction of the state B is stationary, in the long-term simulation.
This suggests the possibility that the state B, unstable in the continuum limit,
may be sustained over a finite period due to the
discreteness in molecules, which forms a domain in space.
The two states A and B coexist in space and form a domain structure.

\begin{table}
\begin{center}
\begin{tabular}{|c|c|c|c|}
\hline
state & $c_{i}$ & continuous & discrete \\
\hline
the unstable fixed point of the R-D eq. &
 $c_{3}=0$ &
 unstable & unstable\\
the stochastic steady state &
 $c_{1}, c_{2}, c_{3} > 0$ &
 (not fixed point) & stable \\
the stable fixed point of the R-D eq. &
 $c_{1}=c_{2}=0$ &
 stable & unstable \\
\hline
\end{tabular}
\end{center}
\caption{Stability of each fixed point
in the continuous reaction-diffusion equation
and in the stochastic system taking into account discreteness in molecules.}
\label{tbl:fp}
\end{table}

\begin{figure}
\begin{center}
\includegraphics[width=80mm]{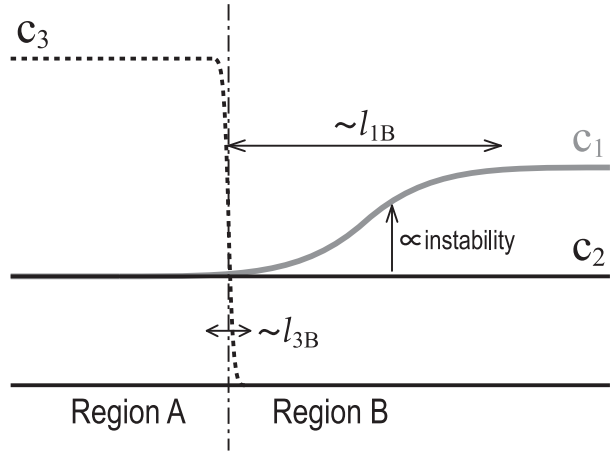}
\end{center}
\caption{Schematic diagram of the border between the regions A and B.
}
\label{fig:sp2-regions}
\end{figure}

First, we consider stability of each of the states in more detail.
The state B is unstable against the inflow of $X_{3}$.
If an $X_{3}$ molecule enters into a region of state B (Region B),
it can be amplified and form a new region (spot) of the state A (Region A).
From linear stability analysis, we find that
the degree of instability of the region B against  the flow of $X_{3}$ is proportional to
$(c_{1} - c_{2})$.
Note that in the region A, $c_{1} \approx c_{2}$, while
in the region B, $c_{1} \approx \sqrt{a} c_{2}$,
and $c_{3}=0$.  Here, the concentration
$c_{2}$ is almost uniform in space because of its long lifetime.
Thus, the degree of the
instability of the Region A,
that is the rate of growth of $c_{3}$, depends
mainly on the distribution of $X_{1}$ (see Fig. \ref{fig:sp2-regions}).

Accordingly, the Kuramoto length of $X_{1}$ is relevant to determine if 
the state B is invaded or not.
Here it should be noted that in the region A, 
$c_{1}$ is smaller than that in the region B.  Hence 
in the vicinity of the region A within the domain of the state B,  $c_1$ is still small, 
as long as it is within the Kuramoto length of $X_{1}$ of the region A.
Thus the instability is weak there,
which prevents a novel region A ($X_{3}$ spot) growing in the vicinity of the existing region A.
Hence, the interval between two neighboring regions A should be longer than 
the Kuramoto length of $X_{1}$.
Assuming that $N_{3} \ll N_{1}$, $N_{2}$ and
$(c_{1}, c_{2}, c_{3}) = (\frac{\sqrt{a}c}{\sqrt{a}+1}, \frac{c}{\sqrt{a}+1}, 0)$ (i.e.,
the unstable fixed point of the reaction-diffusion equation) in the region B,
we obtain the Kuramoto length of $X_{1}$ in the region B as
\[
\ell_{1B} = \sqrt{2D_{1}(k_{4}c_{1})^{-1}} = \sqrt{\frac{200D(\sqrt{a}+1)}{r \sqrt{a}c}}.
\]

Since $X_{3}$ can be amplified by using $X_{1}$ in the region B, 
penetration of $X_{3}$ molecules into the region B must be rare in order
to maintain the region B.
The penetration length is given by the Kuramoto length of 
$X_{3}$, that is computed as
\[
\ell_{3B} = \sqrt{2D_{3}(k_{1}c_{2})^{-1}} = \sqrt{\frac{2D(\sqrt{a}+1)}{br c}}.
\]
for the region B.
For $1 < a \ll (100b)^{2}$, $\ell_{3B} \ll \ell_{1B}$, which implies
that $X_{3}$ molecules seldom reach
the area where $X_{3}$ is strongly amplified.
Thus, the border of regions A and B is maintained
for long time\footnote{For this reason,
we set $D_{3}$ relatively small.
If $D_{3}$ is larger, the border is blurred and the two regions are mixed.}.

On the other hand, due to the fluctuation inherent in SSS, the molecule $X_3$ may be extinct
within some area of the region A,
with some probability.
Hence, the regions A and B
coexist in space, as shown in 
Fig. \ref{fig:sp2-spa-3n2-nr4k-d1}.
As shown, the region A is localized only as spots, and other parts are covered by the region B.

Note that in the corresponding reaction-diffusion equations,
the state A, the stochastic steady state, cannot be realized, while
the state B is unstable.  
Indeed, the reaction-diffusion equation system
is quickly homogenized and converges to the stable
fixed point with $c_{1}=c_{2}=0$.
Hence both the regions A and B, as well as a domain structure from the two,
can exist only as a result of
the discreteness of molecules,
and are immediately destroyed in the continuum limit.

\begin{figure}
\begin{center}
\includegraphics[width=80mm]{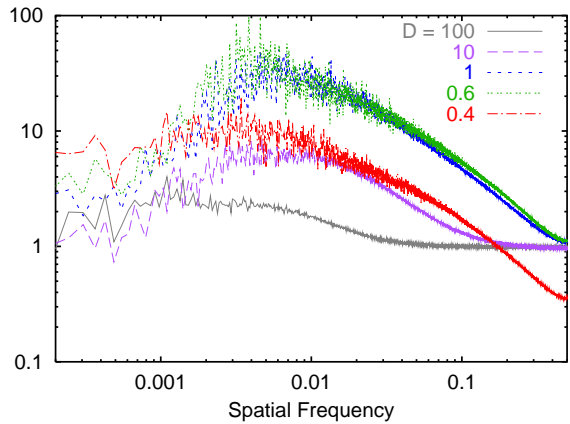}
\end{center}
\caption{Power spectra of the distribution of $X_{3}$.
$r=1$, $a=4$, $b=100$, $N=16384$, and $L_{x}=16384$.
For small $D$, there appears a broad peak at the wave length around the order
of $10^{2}$, which corresponds to the interval of the spots.
For large $D$, $X_{3}$ spreads over the space, and the peak
disappears.}
\label{fig:sp2-powspec}
\end{figure}

Note that in the present model, SSS does not have
so-called the Turing instability\footnote{It is
also possible that the acceleration of reactions
by the discreteness induces or enhances the Turing instability
in certain systems.}, and there is no characteristic wavelength.
Still the spatial structure here has some characteristic length,
as given by the minimal size of the region B,  estimated by the Kuramoto length
$\ell_{1B}$.  In Fig. \ref{fig:sp2-powspec},
we have plotted the spatial power spectrum of the
concentrations of $X_3$.  Although there is no clear peak, there is a
very broad increase around the wavenumber of $0.01$, that corresponds to
average domain size of the region B.

In general, the discreteness of molecules can induce
novel states not seen in the continuum limit.
For example, in a randomly connected catalytic reaction network,
there often exist several fixed points with some chemicals going extinct,
when the number of molecules is small,
while there is only one attractor in the continuum limit.
These discreteness-induced states may coexist in space
in a similar way as discussed above.
Kuramoto length will be a useful index to determine
the behavior around the border of the regions.

\section{Summary and Discussion}

In the present paper, we have reported a novel steady state in 
a system with reaction and diffusion, induced by discreteness in molecules. 
This state cannot be represented by a continuum description,
i.e., partial differential equation (reaction-diffusion equation), but is
sustained by amplification of some reaction due to localization of some molecule $X_j$.
This localization is possible if the molecule species that produce $X_j$
is ``discrete'', in the sense that its average number within the
Kuramoto length of the  $X_j$ molecule is less than 1.  We have formulated a theory
to obtain a self-consistent
solution for the concentrations of $X_i$ and $X_j$,
in relationship with
the amplification rate of the reaction involved with  $X_j$.
For some reaction system, there is a solution with amplification rate larger than 1,
that leads to the existence of stochastic steady state due to the discreteness in molecule number.
The stability of this solution is also computed within this theoretical formulation.

We have also numerically studied a simple reaction-diffusion system
to demonstrate the validity of the theory.
Indeed, a novel stochastic steady state is observed, as predicted theoretically.
We have also extended our theory to include the self-consistent determination of
the Kuramoto length.  Following this extension,
we have provided a numerical  example, to show formation of domains
with different Kuramoto lengths.

The alteration of the steady state by the localization, as well as
our formulation for it is quite general.
Provided that the conditions
\begin{enumerate}
\item[(i)] Chemical $X_i$ generates another chemical species $X_j$.
\item[(ii)] The lifetime of $X_j$ is short or the diffusion of $X_j$ is slow
so that the Kuramoto length of $X_j$ is much smaller
than the distance between $X_{i}$ molecules.
\item[(iii)] The localization of the molecule $X_j$ accelerates some reactions.
\end{enumerate}
Then, the discreteness can alter the dynamics, from that by the continuum description.
The last condition is satisfied if the second or higher order reaction is involved
in the species $X_j$.
Finally, if
\begin{enumerate}
\item[(iv)] the acceleration of the reaction in (iii) alters the density of $X_i$ molecules, through
some reaction(s),
\end{enumerate}
the density of $X_i$ is determined self-consistently with the acceleration factor, 
resulting in a novel steady state.

Note that the localization effect by the discreteness of catalytic molecules itself is also noted by 
Shnerb \textit{et al.} \cite{Solomon2000}.
In their study, however, the density of the catalyst is fixed as an externally
given value. Thus the concentration of the product, localized around the catalyst,
diverges in time.  In our theory, the density of the catalyst ($X_i$)
changes autonomously and reaches a suitable value by following the discreteness effect.

The self-consistent solution scheme to obtain this discreteness-induced stochastic
state can be extended to a case with several components $i_1,i_2,...,i_k$, 
and the corresponding set of chemicals $j_1,j_2,...,j_k$ satisfying 
the conditions (i)--(iii).
In such case, the feedback process in (iv) is not necessarily direct from $j_m$ to $i_m$.
If there is a feedback from the set of chemicals  $j_1,j_2,...,j_k$ to the set
$i_1,i_2,...,i_k$ (condition (iv)$'$), the above scheme 
for the self-consistent dynamics
we presented here works.  With this extension, there is a variety of possibilities, that can
lead to stochastic rhythm or pattern formation induced by discreteness of molecules,
which is not seen in the continuum limit.
For example, in a catalytic reaction network with many components and a limited number
of total molecules,  there always
exist several species that are minority in number, and the conditions (i)--(ii)
are naturally satisfied, while with higher order catalytic reaction the condition (iii) is
often satisfied.
In this case, minority molecules become a key factor to determine a macroscopic state
with rhythm or pattern.
(Note in this case, other molecules can be abundant in number, or indeed it is better to have
such abundant species, so that the stochastic state is stabilized.)

In fact, biochemical reaction networks involve a huge number of species,
while the total number of molecules is not necessarily so large.
In a cell, lots of chemicals work at low concentration
in the order of 1 nM or less.
The diffusion is sometimes restricted, surrounded by macro-molecules,
and may be slow.
In such an environment, it is probable that the average distance
between the molecules of a given chemical species is much larger
than the Kuramoto lengths of some other chemical species.
Some chemicals are localized around some other molecules.
Furthermore, biochemical systems contain various higher order reactions
(for example, catalyzed by enzyme complexes).  In conjunction
with the localization, such reactions can be accelerated.  Hence the conditions (i)--(iii) are
ubiquitously satisfied in intra-cellular biochemical reaction networks.
In addition, since the biochemical reactions involve complex feedback process through
mutual catalytic networks, the condition (iv) or (iv)$'$ is naturally satisfied.

Accordingly, it will be important to study the amplification of
some reaction and its maintenance through feedback will be relevant to
biochemical reactions.  Indeed,
some molecules that are minority in number sometimes
play a key role in biological function.
Relevance of minority molecules is also  
discussed from the viewpoint on a control mechanism of a cell, in relationship
with the kinetic origin of information \cite{minorityJTB,minoritynetworkinPRE}.

The importance of our theory is not restricted to biological problems.
Verification of our
result will be possible by
suitably designing a reaction system, with the use of, say,
microreactors or vesicles.  The acceleration and maintenance of some reactions
by localization of molecules will be important to design some function 
in such micro-reactor systems.


\paragraph{Acknowledgement}

The present paper is dedicated to Professor Yoshiki Kuramoto
on the occasion of his retirement from Kyoto University.
With the papers \cite{Kuramoto1,Kuramoto2} that introduced
Kuramoto length a novel research field was opened; the study
of chemical wave and turbulence with the use of {\sl continuous, deterministic}
reaction-diffusion equation.  It is our pleasure to use his length in the opposite
context here, for the description of novel steady
states in {\sl discrete, stochastic}  reaction-diffusion systems.
The present work is supported  
by grant-in-aid for scientific research from the Ministry of Education,
Culture, Sports, Science and Technology of Japan (15-11161),
and the Japan Society for the Promotion of Science.

\end{document}